# BER PERFORMANCE ANALYSIS FOR WIMAX PHY LAYER UNDER DIFFERENT CHANNEL CONDITIONS:

Shantanu Pathak[1] and Ranjani S.[2]

Department of Telecommunication, SRM University, Chennai, India

shantanupathak614@gmail.com , ranjani.s@ktr.srmuniv.ac.in

## ABSTRACT

*This paper gives an introduction on the IEEE 802.16 standard – WIMAX or Worldwide Interoperability for Microwave Access. The different parts give details on the architectural specifications of WiMAX networks and also on the working principle of WiMAX networks including its services provided. It also provides brief descriptions on its salient features of this technology and how it benefits the networking industry. A brief outline of the basic building blocks or equipment of WiMAX architecture is also provided. This paper also evaluates the simulation performance of IEEE 802.16 OFDM PHY layer. The Stanford University Interim (SUI) channel model under varying parameters is selected for the wireless channel in the simulation. The performance measurements and analysis was done in simulation developed in MATLAB.*

## KEYWORDS

*WiMAX, IEEE, 802.16, OFDM, MIMO, FEC, SUI, PHY, CP*

## 1. INTRODUCTION:

WiMAX is a digital broadband wireless access system; derived from the IEEE 802.16 standard which has been designed for wireless metropolitan area networks and hence also known as WirelessMAN. WiMAX systems can provide broadband wireless access (BWA) up to 50 km (30 miles) for fixed stations, and 5 - 15 km (3 – 10 miles), for mobile stations. It is maintained by the WiMAX Forum industry alliance; that are responsible for the promotion and certification, compatibility and interoperability of products based on the IEEE 802.16 standards.

The main development that WiMAX has brought in the field of telecommunication is that it is able to provide broadband internet access in a wireless manner. The present broadband internet access generally includes access through wired access or through Wi-Fi access. However providing wired access in a larger scale becomes very expensive and the range and security of Wi-Fi is not reliable. Hence the new technology WiMAX was developed to provide hi-speed internet access in a wireless manner instead – wireless because it would much less expensive then wired; and much more broad coverage than normal Wi-Fi, like cell phone networks, so that it is able to extend to much larger areas.

## 2. How WiMAX Works:

In general, WiMAX operates similarly to Wi-Fi but with longer range, higher speeds and operating larger number of users. WiMAX could potentially erase the suburban and rural blackout areas that currently have no broadband Internet access because phone and cable companies have not yet run the necessary wires to those remote locations.

A WiMAX system consists of two parts: a WiMAX Base Station and a WiMAX Receiver.

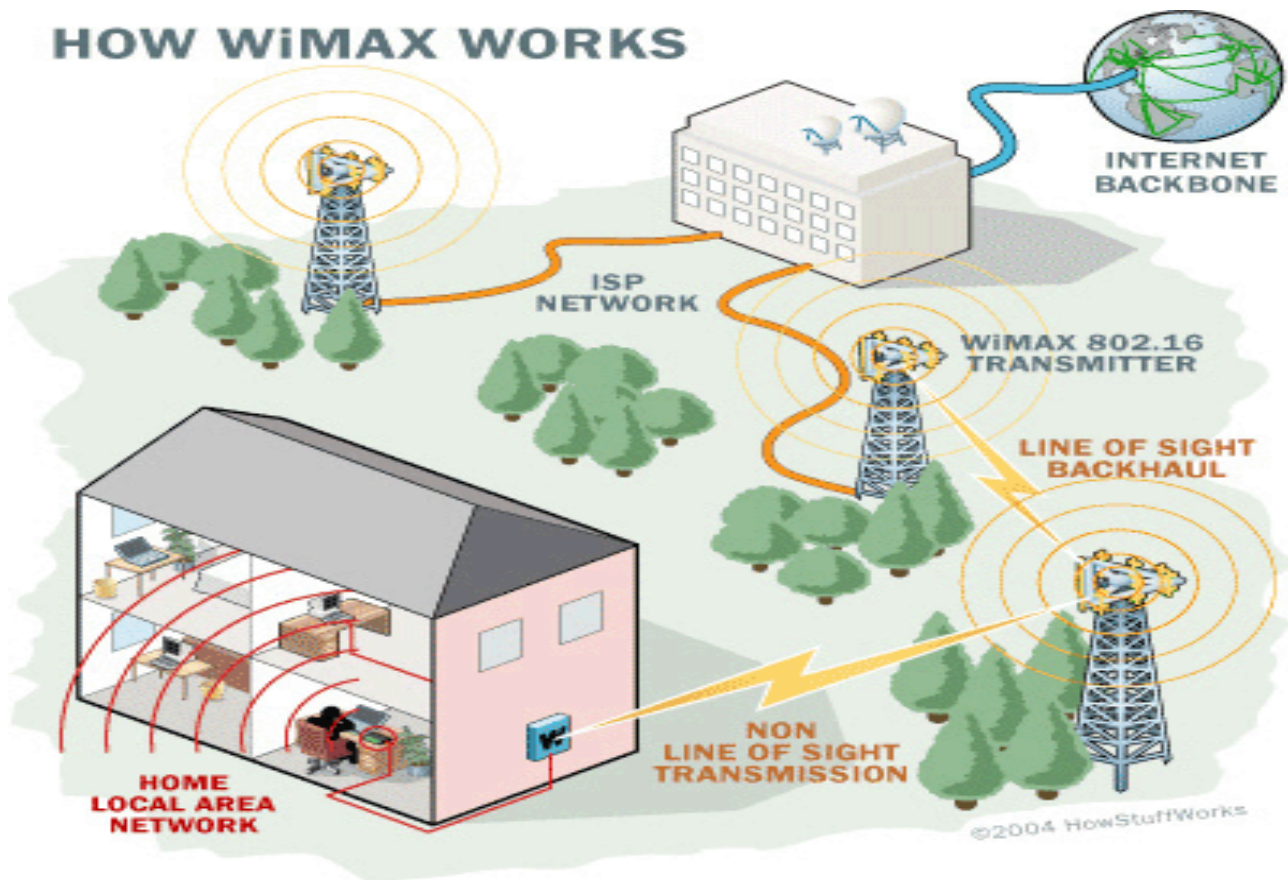

## 2.1 WiMAX Base Station:

Base station is place where WiMAX signals are broadcasted and it basically consists of electronic devices and WiMAX towers. WiMAX towers are further categorised as fixed and mobile towers.

The fixed towers which are connected with home or office subscriber units, provided indoor or outdoor subscriber unit operate on high power output transmissions providing 3.5 GHz signals out to distance of 25 miles when passing through WiMAX fixed towers arranged in mesh topology.

Mobile towers accommodate smaller devices such as laptops with WiMAX wireless mini cards, cellular phones and mobile devices. These mobile towers work entirely different from fixed towers due to hand held devices being limited to battery and processing power.

## 2.2 WiMAX Receiver:

WiMAX receiver is also termed as CPE or customer premises equipment. CPE is a device which receives the signal from base station and connects to the WiMAX networks. These devices are usually a Wireless USB modem or PCMCIA slot card for laptops or computers. WiMAX subscriber units can be indoor or outdoor depending on the distance to the nearest WiMAX base station tower.

A WiMAX tower station can connect directly to the Internet using a high-bandwidth, wired connection. It can also connect to another WiMAX tower using a line-of-sight, microwave link. This connection to a second tower referred to as a backhaul, along with the ability of a single tower to cover up to 3,000 square miles allows WiMAX to provide coverage to remote rural areas.

WiMAX can provide two types of services:

*Non Line of sight service:*

In this mode a small antenna connects to the tower. WiMAX uses a lower frequency range of 2 GHz to 11 GHz (similar to Wi-Fi). Lower-wavelength transmissions are not as easily affected by physical obstructions. They are better able to diffract, or bend, around obstacles.

*Line of sight service:*

In this mode a fixed dish antenna points straight at the WiMAX tower from a rooftop or pole. The line-of-sight connection is stronger and more stable, so it's able to send a lot of data with fewer errors. Line-of-sight transmissions use higher frequencies, with ranges reaching a possible 66 GHz. At higher frequencies, there is less interference and lots more bandwidth.

## 3. WiMAX Architecture:

The IEEE 802.16e-2005 standard provides the air interface for WiMAX but does not give proper specifications for the WiMAX network. The WiMAX Forum's Network Working Group (NWG) is responsible for developing the network requirements, architecture, and protocols for WiMAX, using IEEE 802.16e-2005 as the air interface.

The overall WiMAX Architecture can be logically divided into three parts:

1. Mobile Station (MS): It is used by the subscriber or end user as a source of network connection. The base station (BS) provides the air interface for the mobile stations.

2. Access Service Network (ASN): It contains ASN gateways to build the radio access at the end and comprises of more than two or three base stations. ASN is responsible for radio resource management, encryption keys, routing to the selected network and client functionality.

3. Connectivity Service Network (CSN): It is responsible for providing IP functions. It is responsible for internet connections, corporate and public networks and many other user services.

There are two scenarios for a wireless deployment:

*Point-to-point (P2P):*

P2P is used when there are two points of interest- one sender and one receiver. P2P can also be used as a point for distribution using point-to-multipoint (PMP) architecture. The outturn of P2P radios will be higher than that of PMP products. It is also known as fixed WiMAX designated under IEEE 802.16 -2004 standard.

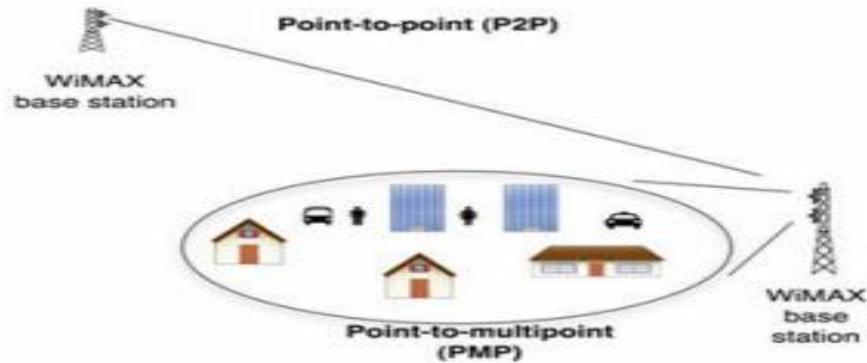

Point-to point and point-to-multipoint configurations

*Point- to-Multipoint (PMP):*

Point to Multipoint is used for broadcasting or distribution services. One base station can service hundreds of different subscriber stations in terms of bandwidth and services offered. It is also known as mobile or wireless WiMAX designated under IEEE 802.16e-2005 standard.

## 4. WiMAX Equipment:

WiMAX architecture is built upon two components: radios and antennas. Most WiMAX systems are designed such that there is a base station radio separate from the antenna. The antenna is connected to WiMAX radio via a cable known as a "pigtail".

### 4.1 WiMAX Radios:

A WiMAX radio contains both a transmitter and a receiver to simultaneously send and receive. It generates signals at a frequency known as the carrier frequency (usually between 2 and 11 GHz).

### 4.2 WiMAX Antennas:

WiMAX antenna, are designed to provide efficient performance for the required application. There are three man types:

*Omni-directional:* Omni directional antennas are used for point-to-multipoint configurations.

*Sector:* A sector antenna focuses the beam in a specific focused area, offers greater range and throughput with less energy.

*Flat-Panel:* Panel antennas are most often used for point-to-point applications.

## 5. WiMAX Features:

### 5.1 WiMAX OFDM-based air interface (PHY Layer):

The WiMAX air interface (PHY Layer) is based on Orthogonal Frequency Division Multiplexing or OFDM. The WiMAX RF signals use OFDM techniques and its signal bandwidth can range from 1.25 to 20 MHz. To maintain orthogonality between the individual carriers the symbol period must be reciprocal of the carrier spacing.

OFDM is based on a transmission scheme called multi-carrier modulation , which divides a high bit stream into a number of low bit streams, which are each modulated by separate carriers called subcarriers or tones.

These modulation schemes generally remove Inter Symbol Interference by keeping the symbol duration large. But for high data rate systems where the symbol duration is small, the high bit stream is splitted into several parallel streams increasing the symbol duration of each stream. The sub-carriers are chosen such that they are orthogonal; hence there is no mutual interference among the signal streams. This helps in removing ISI to some extent; but to completely remove it guard bands are also added in the OFDM scheme.

### 5.2 Adaptive Antenna Techniques:

Adaptive Antenna Systems (AAS) used in WiMAX utilizes beam-forming techniques for focusing and directing the wireless beam/signal between the base station and the receiver station. This reduces interference from other external sources and noises, as the beam is focused directly between two points. Dynamic Frequency Selection (DFS) schemes are applied in which air waves are first scanned to determine where interference doesn't occur, and specific frequencies in that area where no interference occurs is selected.

WiMAX uses these techniques through the use of MIMO (multiple-input multiple output) communication schemes. MIMO uses multiple antennas at both transmitter and receiver, which sets up a multiple data stream on the same channel, increasing the data capacity of the channel to a great extent. With multiple antennas at both transmitter and receiver, the transmitter and receiver can synchronise and coordinate to move to a free frequency band if and when interference occurs. The use of these techniques provides lots of advantages on the basis of coverage, self-installation, power consumption, frequency re-use and bandwidth efficiency. Use of beam form techniques and MIMO under WiMAX reduces interference while increasing throughput and efficiency.

### 5.3 Adaptive Modulation and Coding:

WiMAX supports a variety of modulation and coding schemes in its specifications. These schemes are adaptive, that is they can be made to vary the parameters, according to the prevalent conditions; they can also be made to change on a burst-by-burst basis per connection.

The coding and modulation schemes maintain a steady signal strength by using different schemes over increasing distance ; this is achieved by decreasing throughput over range so as to provide the best Quality of Service (QoS) possible over distance. Also, a technique called Dynamic Bandwidth Allocation (DBA) is used, which monitors the network for interference or reduction in signal strength, the base station allocates more bandwidth to the afflicted stream.

To determine the required WiMAX modulation and coding scheme the channel quality feedback indicator is used. The mobile (UE) can provide the base station (BS) with feedback on the downlink channel quality and for the uplink, the base station can estimate the channel quality, based on the received signal quality.

The various modulation schemes used under WiMAX:

*Downlink*:   BPSK, QPSK, 16 QAM, 64 QAM; BPSK optional for OFDM-PHY.

*Uplink*: BPSK, QPSK, 16 QAM; 64 QAM optional.

The various coding schemes used under WiMAX:

*Downlink*: convolutional codes at 1/2, 2/3, 3/4, 5/6

*Uplink*: convolutional codes at 1/2, 2/3, 3/4, 5/6

## 6. Simulation Model:

The design of the simulated transmitter and receiver for the WiMAX OFDM based PHY layer is shown below. This design is based on the physical layer of the IEEE 802.16 2004 WIMAX OFDM air interface. In this setup, we have used for simulation only the compulsory features of the specification, while omitting the other features for future improvements. Channel coding part is divided of three parts Randomization, Forward Error Correction (FEC) and Interleaving.

FEC is done in two steps, firstly through Reed Solomon (RS) Code and then using Convolutional Code (CC). These are applied in the reverse order at channel decoding at the receiver side. The complete channel encoding and decoding setup design is shown.

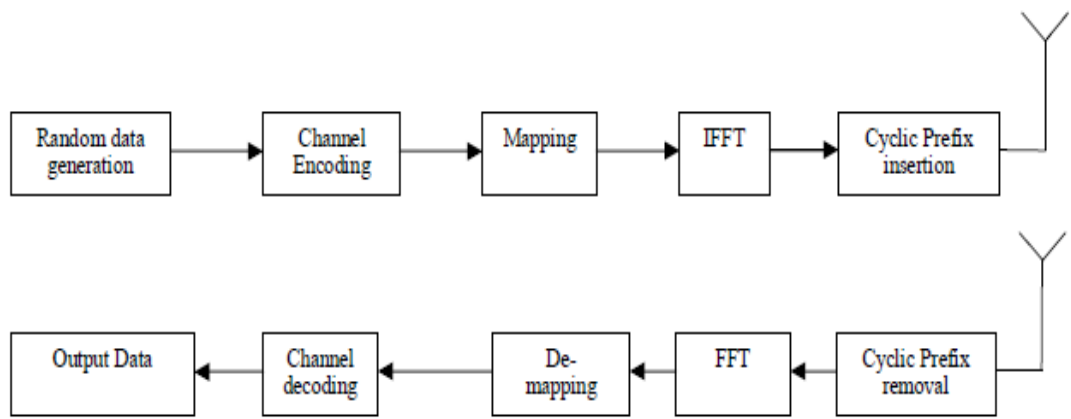

Simulation Setup

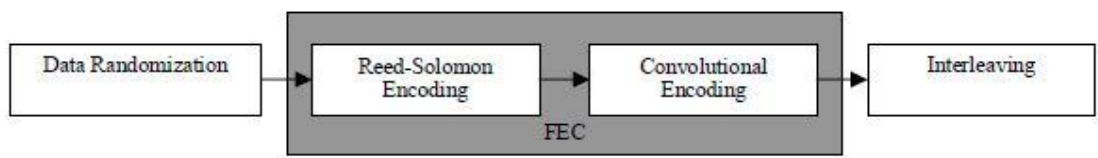

Channel Encoding Setup

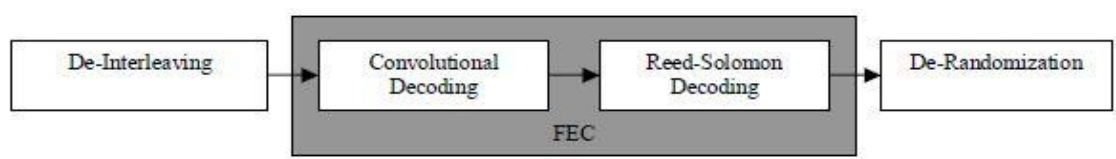

Channel Decoding Setup

This structure is based on the physical layer of the IEEE 802.16 Wireless MAN OFDM air interface. The above given structure is explained in detail.

## 6.1 Randomization:

A Pseudo Random Binary Sequence generator is used to implement randomization. It uses a 15-stage shift register with a generator polynomial of $1+x^{14}+x^{15}$ with XOR gates in feedback configuration.

## 6.2 Forward Error Correction:

This block consists of two layers: a first layer of Reed-Solomon coding followed by a second layer of Convolutional coding. Reed Solomon codes are block codes that are quite good in correcting burst errors, while convolutional codes are efficient in correcting random errors generated during transmission in multipath channels. Both of them together provide an effective measure to correct most type of errors generated in the channels.

### 6.2.1 Reed Solomon Encoder:

A Reed-Solomon code is defined as RS (N, K, and T) with l-bit symbols. The RS(N,K, and T) means that for encoding, the encoder uses K data symbols of l bits each and adds 2T parity symbols to create an N-symbol code word. These N, K and T can be defined as:

- N: total no. of bytes after the encoding process,
- K: total no. of data bytes before the encoding process,
- T: total no. of data bytes that can be corrected.

The error correction capabilities of a given RS code is given by (N – K), which is defined as the measure of redundancy in the block. If the position of errors in symbols is not known, a Reed-Solomon code can used to correct up to T symbols, where T can be defined as T= (N – K)/2.

As per the standard, the RS code encoding has been created from a systematic RS (with N =255, K = 239, and T= 8) code, which uses a Galois field given as GF $(2^8)$.

The given polynomials are used for code generator and field generator:

$G(x) = (X+\lambda^0)(X+\lambda^1)(X+\lambda^2)\ldots(X+\lambda^{2T-1})$, $\lambda = 02_{HEX}$

$P(x) = X^8+X^4+X^3+X^2+1$.

### 6.2.2 Convolutional Encoder:

After the RS coding, the encoded data is passed through a convolutional encoder. The encoder has native rate of 1/2, a constraint length of 7 and the given generator polynomials are used to produce its two code bits. The generator is shown below.

$G1 = 171_{OCT}$   ; For X output

$G2 = 133_{OCT}$   ; For Y output

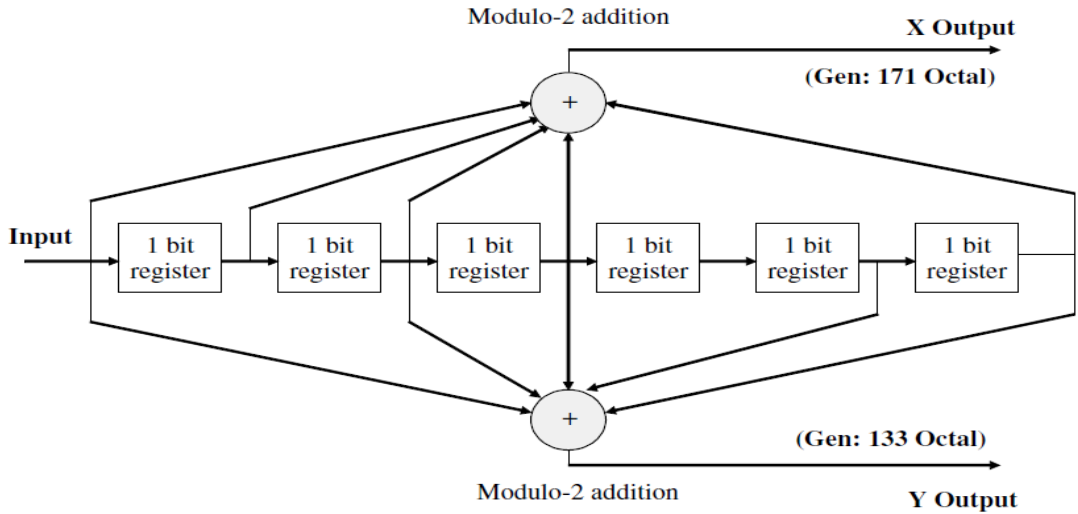

Rate 1/2 convolutional encoder with memory 6

In order to get different coding rates, a process knows as a puncturing operation is performed on the output of the convolutional encoder. A punctured convolutional code that is generated from the mother/native code rate 1/2, is performed with generator polynomials.

For decoding the received data, a soft input maximum likelihood sequence estimator utilized through the Viterbi algorithm is used. The block sizes and code rates used in the modulations in WiMAX systems as specified in IEEE 802.16 standard are shown.

| modulation | Uncoded block size (bytes) | Coded block size (bytes) | Overall code rate | RS code | CC code rate |
|---|---|---|---|---|---|
| BPSK | 12 | 24 | 1/2 | (12,12,0) | 1/2 |
| QPSK | 24 | 48 | 1/2 | (32,24,4) | 2/3 |
| QPSK | 36 | 48 | 3/4 | (40,36,2) | 5/6 |
| 16-QAM | 48 | 96 | 1/2 | (64,48,8) | 2/3 |
| 16-QAM | 72 | 96 | 3/4 | (80,72,4) | 5/6 |
| 64-QAM | 96 | 144 | 2/3 | (108,96,6) | 3/4 |
| 64-QAM | 108 | 144 | 3/4 | (120,108,6) | 5/6 |

Mandatory channel coding per modulation in WiMAX

### 6.3 Interleaver:

A block interleaver then interleaves the generated RS-CC encoded data. Under the standard IEEE 802.16, the interleaver is defined in two steps of permutations. The first results in adjacent coded bits being mapped onto nonadjacent subcarriers. The second permutation results in adjacent coded bits are mapped alternately onto less or more significant bits of the constellation, thus reducing long streams of unusable bits.

### 6.4 Data Symbols Modulation Mapper:

The bit interleaved data are then passed through the constellation mapper, where depending upon the size of the input interleaved data, modulated data is generated using one of the following four different modulation schemes: BPSK, QPSK, 16-QAM and 64-QAM.

### 6.4.1 Subcarriers Allocation:

The WiMAX standard specifies three types of subcarriers, which are known as, data, pilot and null subcarriers. The mapped data are to be arranged in a matrix where the rows number is equal to the number of data subcarriers. The pilot and null subcarriers are then inserted to form the total subcarriers. Table below gives classification of subcarriers for both fixed and mobile WiMAX systems according to the IEEE 802.16d and 16e standards.

| Subcarriers type | Fixed WiMAX | Mobile WiMAX |
|---|---|---|
| $N_{data}$ | 192 | 360 |
| $N_{pilot}$ | 8 | 60 |
| $N_{guard}$ | 56 | 92 |
| $N_C$ | 256 | 512 |

### 6.5 IFFT:

The IFFT is used to generate a time domain signal, for the symbols generated after modulation can be considered as the amplitudes of a certain range of sinusoids. Each of the discrete samples before applying the IFFT algorithm is for an individual subcarrier. Besides ensuring the orthogonality of the OFDM subcarriers, the IFFT represents also a quick and efficient way for modulation of the subcarriers in parallel, and hence, the reducing the use of multiple modulators and demodulators.

The generated data after subcarrier allocation are provided to the IFFT for time domain mapping.

### 6.6 Cyclic Prefix Insertion:

To reduce the effects of multipath, a cyclic prefix is added to the time domain samples. Four different cyclic prefix types/durations are provided in the standard. It is taken as G, which is defined as the ratio of CP time to OFDM symbol time, and can be equal to 1/32, 1/6, 1/8 and 1/4.

### 6.7 Stanford University Interim (SUI) Channel Model:

SUI or, the Stanford University Interim Channel Model is a set of 6 channel models for three different terrain types and also provide a variety of Doppler spreads, delay spreads and line-of-sight/non-line-of-site conditions (LOS/NLOS). It defines six channels to address three different terrain types.

These three terrain types are defined as A, B and C; Terrain A is hilly terrain with moderate to heavy tree density and high path loss; Terrain B is hilly terrain with light tree density or flat terrain with moderate to heavy tree density and moderate path loss; and C is mostly flat terrain with light tree density and has low path loss.

These models can be used for simulation, testing, and design of technologies used for fixed BWA applications.

| Channel | Terrain Type | Doppler Spread | Spread | LOS |
|---------|--------------|----------------|--------|-----|
| SUI-1 | C | Low | Low | High |
| SUI-2 | C | Low | Low | High |
| SUI-3 | B | Low | Low | Low |
| SUI-4 | B | High | Moderate | Low |
| SUI-5 | A | Low | High | Low |
| SUI-6 | A | High | High | Low |

SUI Channels and their characteristic parameters

After the wireless channel we have the receiver model, the receiver basically performs the reverse operation of the transmitter as well as channel estimations which are required to generate the unknown channel coefficients.

Firstly, the CP is removed and the received signal is converted to the frequency domain using, the FFT algorithm. An OFDM symbol is composed by data, pilots, and null sub carriers. Hence a process to separate and divide all the subcarriers is necessary in between. Firstly the guard bands are removed followed by a disassembling, which is done to generate pilots and data subcarriers. The pilot subcarriers are used in channel estimation, for calculating the channel coefficients. To perform an equalization of the data these estimated channel coefficients can be used in the demapper, so as to compensate for the frequency selective fading of the multipath propagation channel. After demapping of data, it enters the decoder block.

## 7. Simulation Results:

### 7.1 Physical layer performance results:

The aim behind simulating the physical layer of WiMAX was to study BER performance under different channel conditions and varying parameters, which are used to determine and characterise the performance of these systems.

### 7.2 Performance Analysis of WiMAX PHY layer with and without channel coding:

Here simulation of FEC, that is, the combined Reed Solomon and Convolutional code; are done to check how much performance reduction will be created in the system (if such codes are not utilised in the system ). To test this simulation, SUI-3 channel model are used.

We define the following input parameters in our simulations:

- Length of CP: 1/4, 1/8, 1/16, 1/32
- Type of Modulation Used: BPSK, QPSK, 16QAM or 64QAM
- Channel type: SUI (1-6)
- Channel Bandwidth: 1.25 – 28 MHZ

In this section we have presented various BER vs. SNR plots for the different modulations under SUI –3 Channel Model.

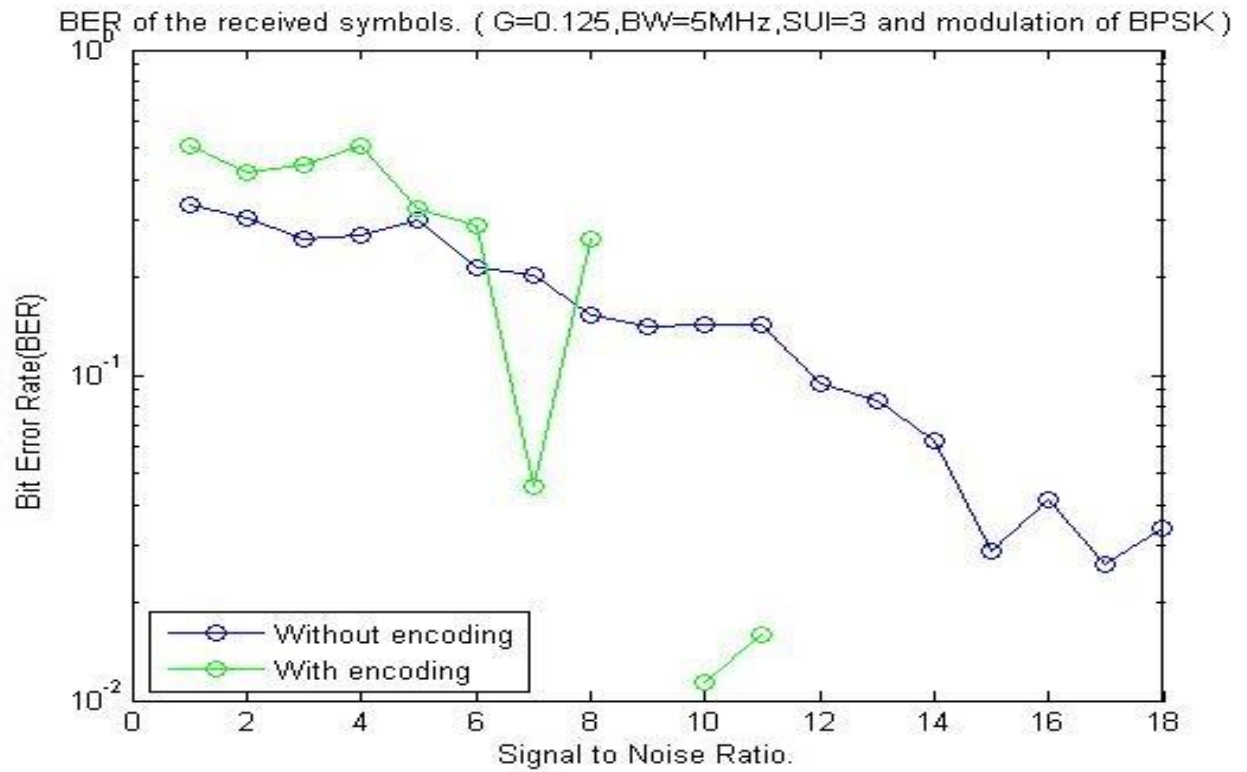

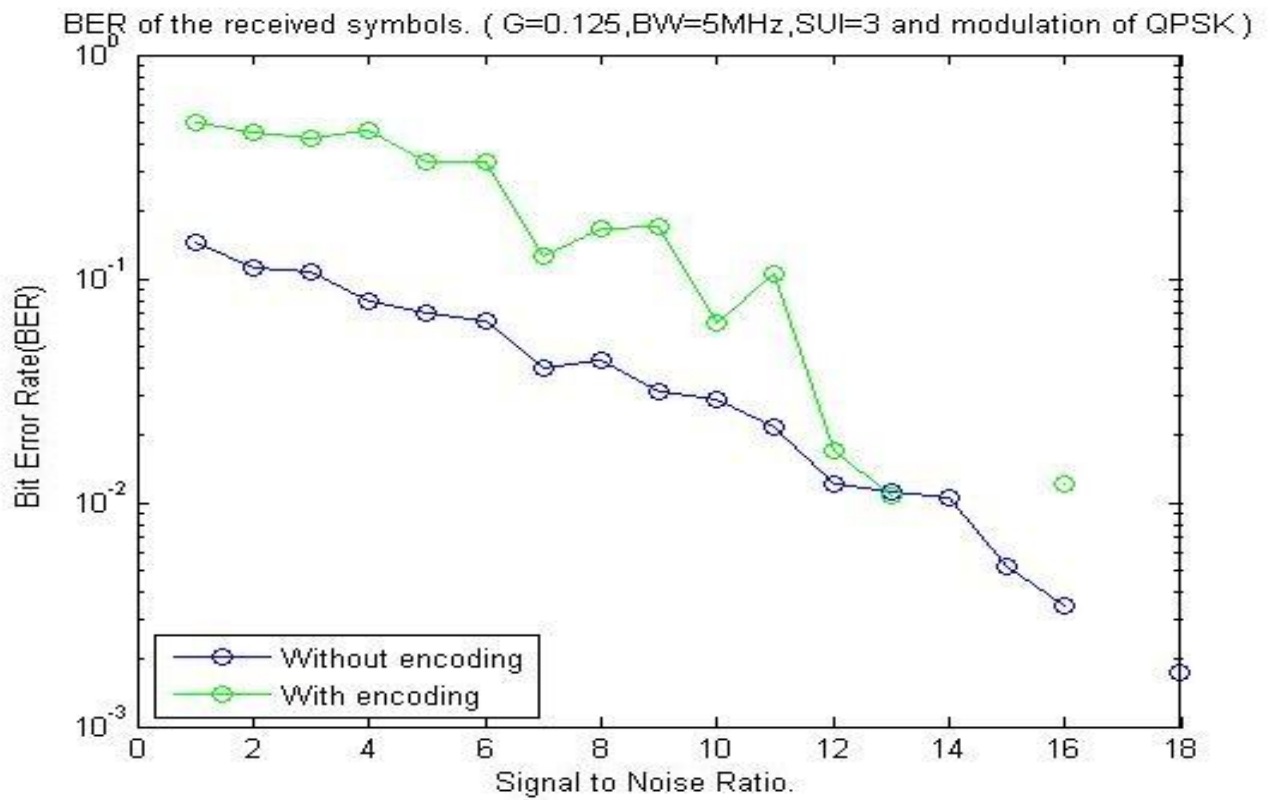

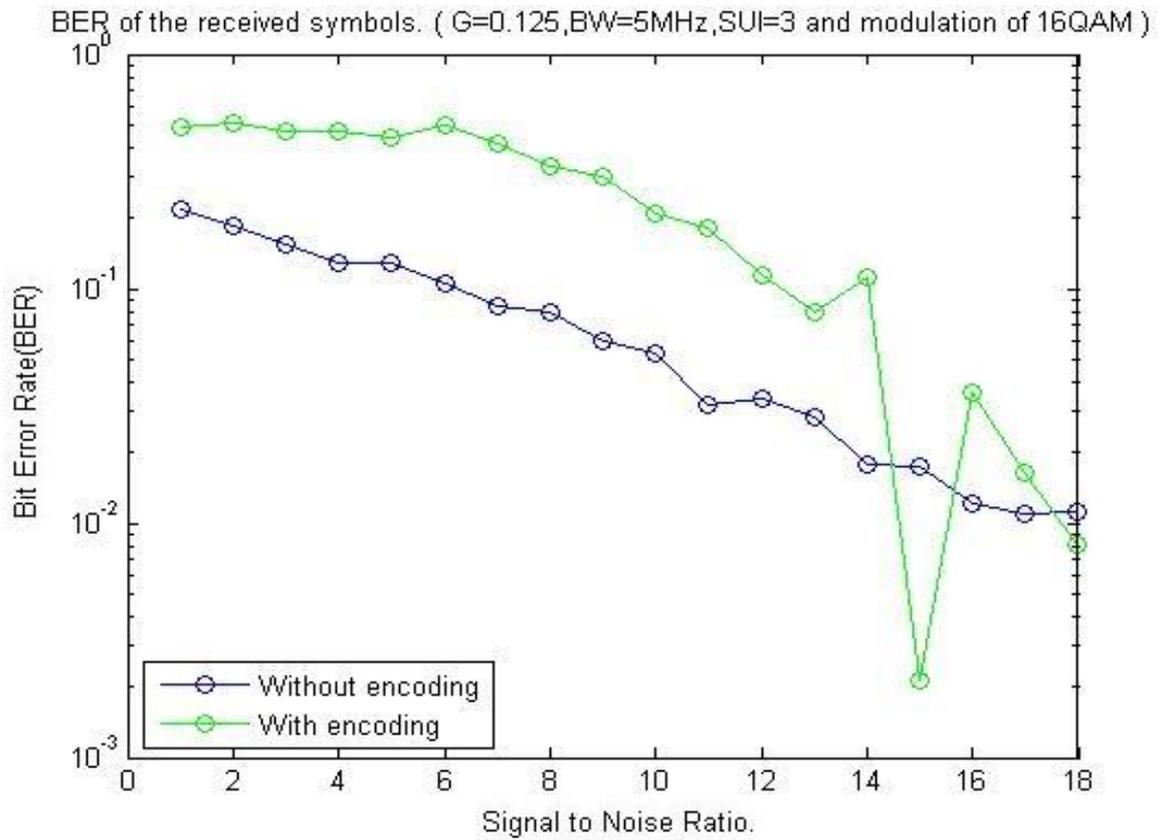

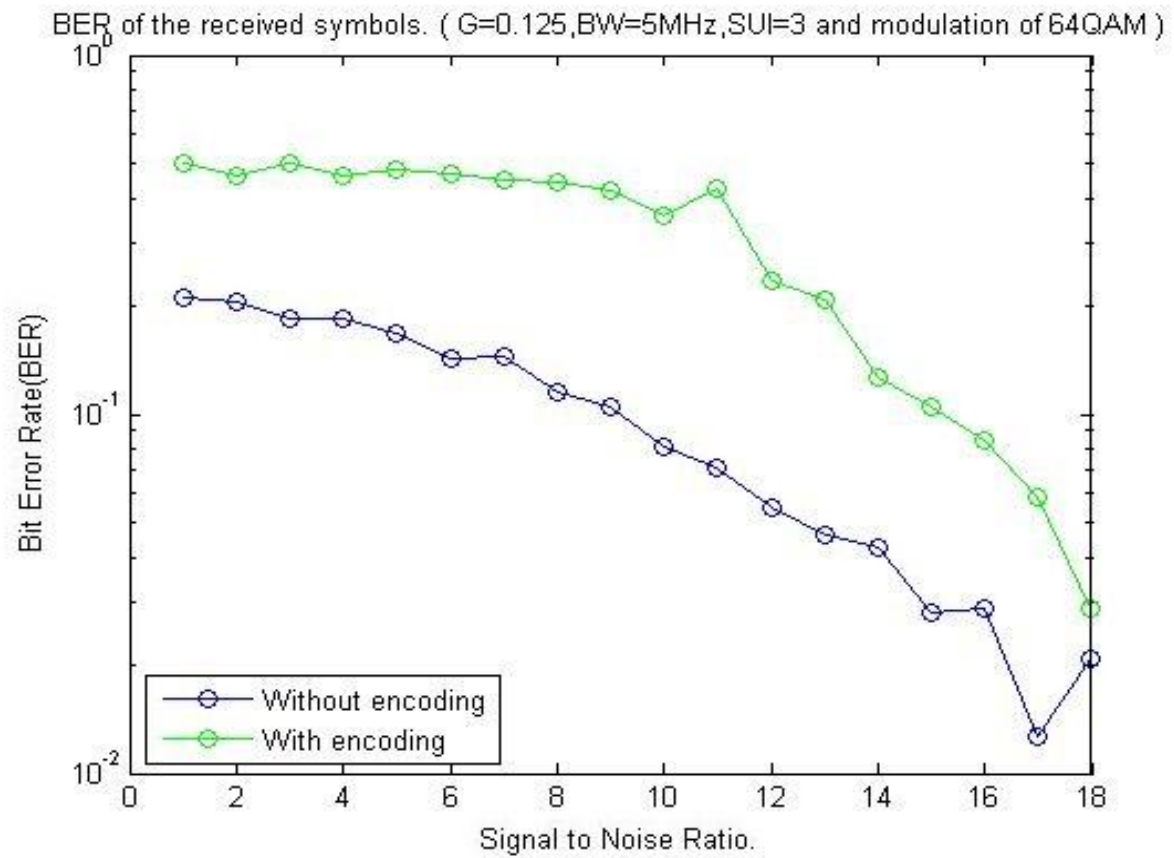

The chief advantage of cyclic codes is that they provide much ease in encoding and also provide a well-defined mathematical structure for efficient decoding schemes. Reed Solomon codes are block codes that are quite good in correcting burst errors, while convolutional codes are efficient in correcting random errors generated during transmission in multipath channels. Both of them together are quite efficient in correcting most type of errors generated in the wireless channels.

Reed-Solomon codes are one most powerful and efficient code to correct the errors generated during transmission, and hence by using this codes in WIMAX we can reduce the bit-error in noisy environment and useful to provide the efficient data to the subscriber. And by using RS-CC codes, further reduction of bit error is achieved. Thus utilizing FEC we can transmit very high data rates in limited bandwidth channels.

**7.3 Performance analysis of WiMAX PHY layer using different values of CP:**

In the simulation model of WiMAX PHY Layer, data is modulated and then Cyclic Prefix is added to it to reduce the effect of fading, and also for receiver synchronisation. In this model SUI – 3 channel is considered.

Comparing transmitted data and demodulated data, we get the Bit Error Rate.

We define the following input parameters in our simulations:

- Type of Modulation Used: BPSK, QPSK, 16QAM or 64QAM
- Channel type: SUI ( 1-6)
- Channel Bandwidth: 1.25 – 28 MHZ

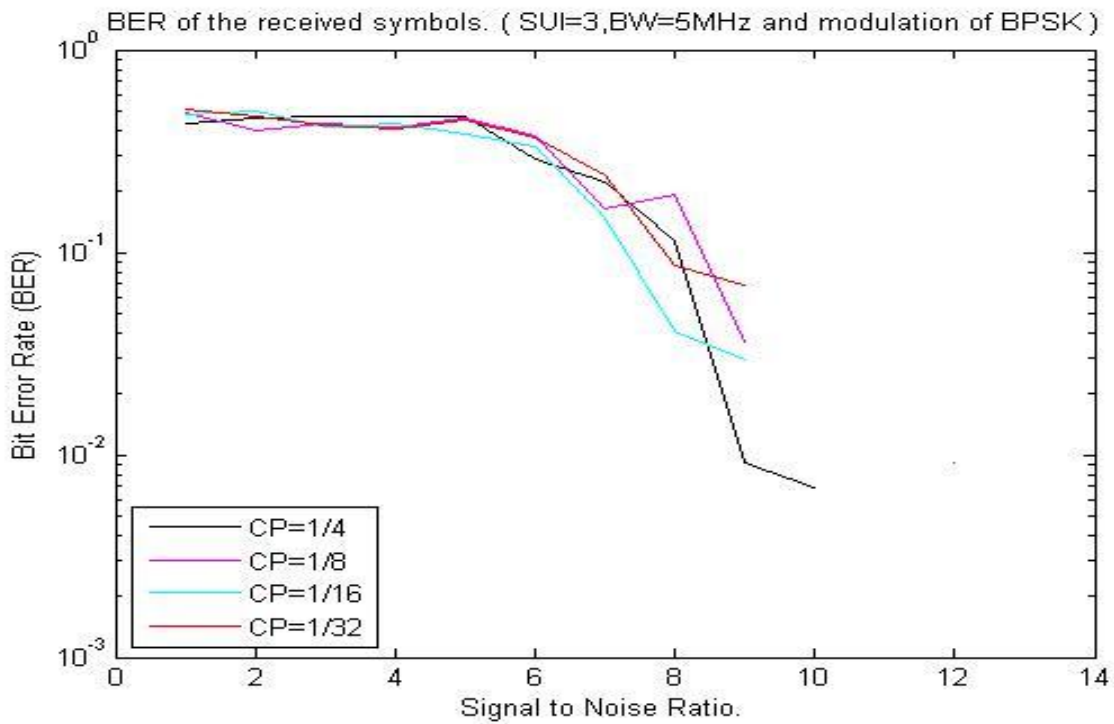

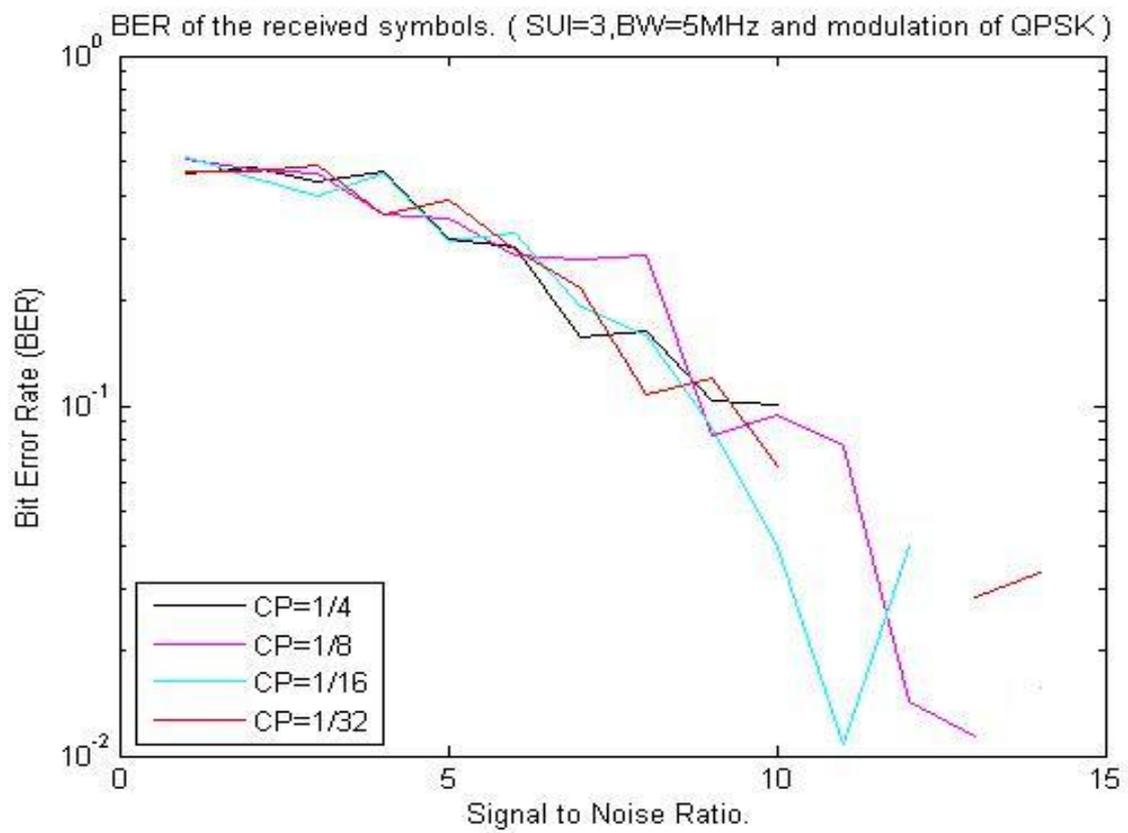

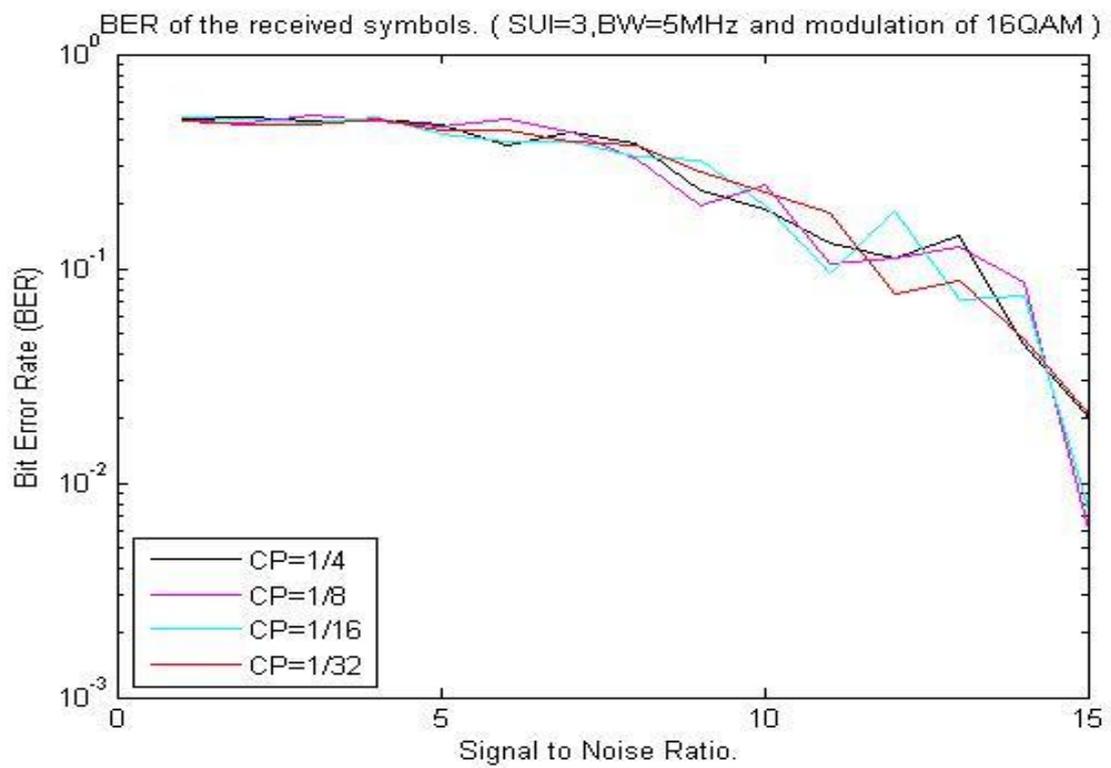

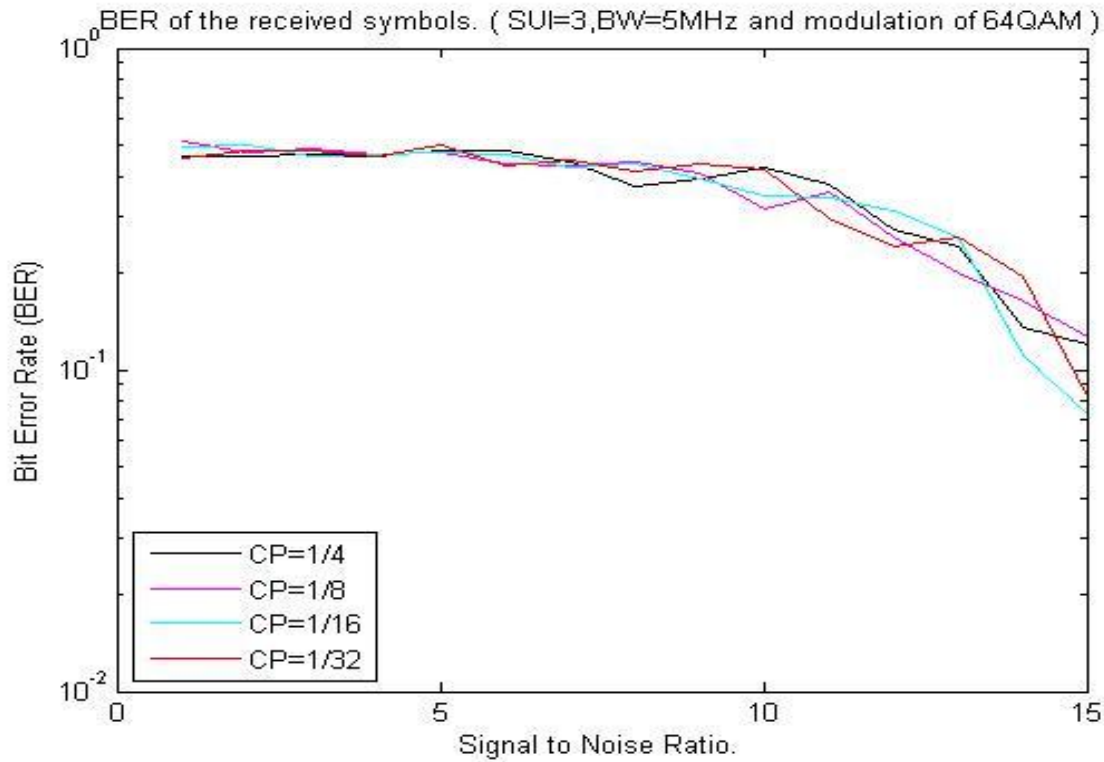

CP is quite an important factor in WiMAX systems. CP reduces the speed and rates of transmission of data, but it is also required to combat the effects of multipath propagation. Hence large CP generally used for reducing effects of multipath. Large CP also means large time gaps between two frames, hence giving extra time to receiver to receive signals from multipath channels. Though large CP has a disadvantage of reducing data rates, it increases coverage area to long distances, and hence providing choice of CP as per coverage area requirements.

With increasing distance, signal strength as well as SNR decreases; as a solution to decreasing CP we use lower modulation levels from higher modulation levels. At low SNR, fading is more and signal strength is going low as distance increases.

Thus depending upon the requirements of the system, the CP value is to be defined and used for in the OFDM symbol.

**7.4 Performance Analysis of WiMAX PHY layer under Different SUI Channel Models:**

Here we simulate and analyse the performance of the WiMAX systems by varying the modulation techniques in different SUI channel models.

Here we have simulated the WiMAX PHY layer under the different SUI channel models and analyse their BER vs. SNR plots

We provide the following input parameters:

- Length of CP : 1/4, 1/8, 1/16, 1/32
- Type of Modulation Used : BPSK, QPSK, 16QAM or 64QAM
- Channel Bandwidth : 1.25 – 28 MHZ

In this section, various BER vs. SNR plots presented which obtained for all the essential modulation profiles in the standard with the same ratio of cyclic prefix and almost equal channel bandwidth.

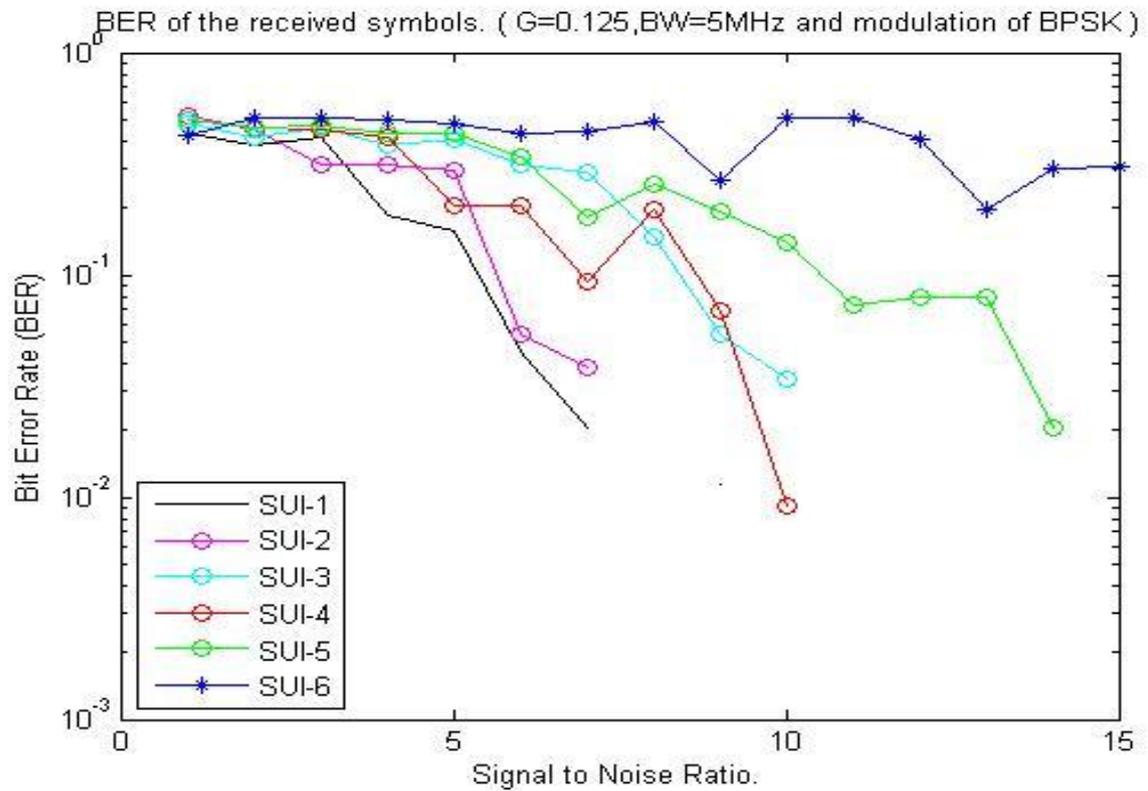

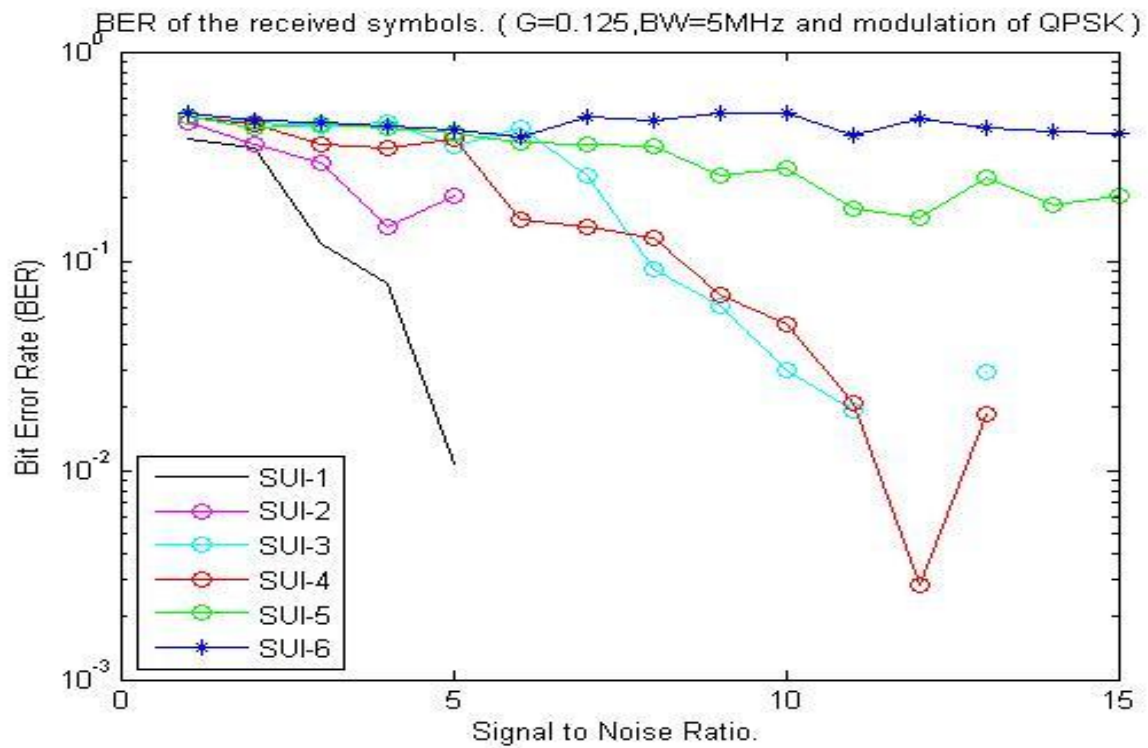

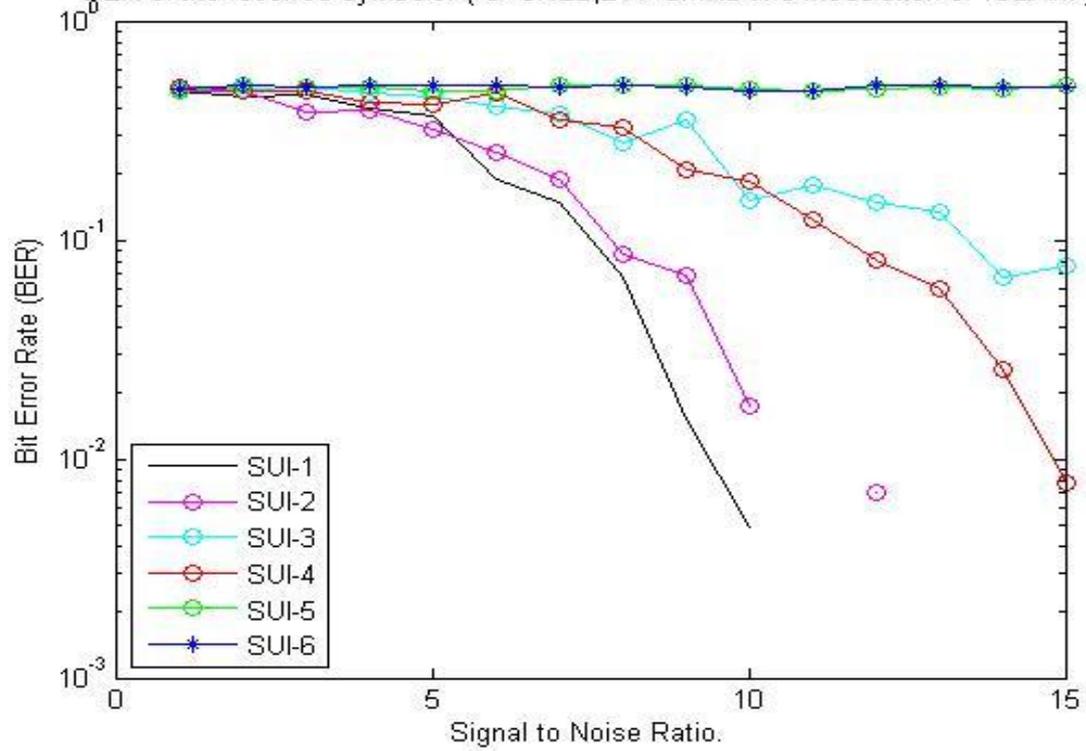
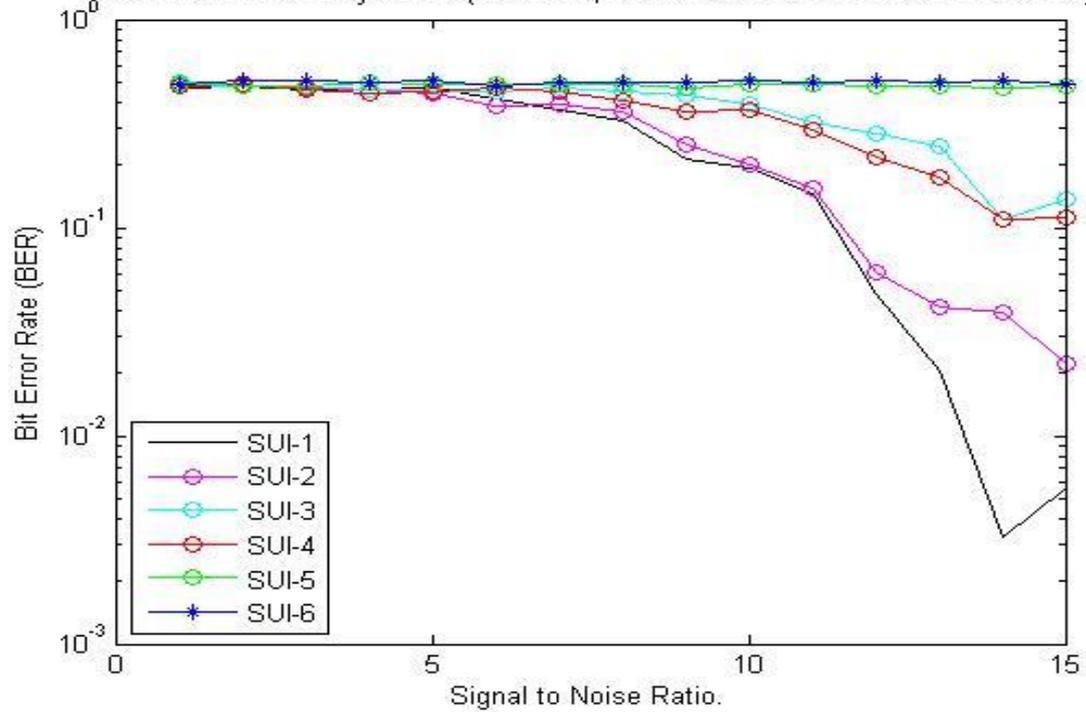

Here the performance analysis of Fixed WIMAX PHY layer for OFDM on different channel condition has been done.

The channels in Category C (SUI1 SUI2) show minimum path loss, due to their flat terrain with light tree densities.

The channels in Category B (SUI3 SUI4) provide results between Category A and Category B, showing moderate path loss (due to terrain being a mixture of hills with light tree density and plains with moderate tree density)

The channels in category A (SUI5 SUI6) contain more noise and suffer from high attenuations (due to terrain of hills with heavy tree density). Hence higher SNR values are required, thus requiring more power for bits to be transmitted over a high fading channel from base station to receiver station.

## 8 Conclusions:

In this paper we provide an introduction to the WiMAX standard and its specifications and parameters. This paper also gives on a simulation of the WiMAX PHY layer under various channel conditions. The simulated PHY layer supports all the modulation and coding schemes defined in the standard, as well as the different CP lengths. At the receiver, perfect channel estimation is assumed. The performance evaluation was done, on the basis of performance analysis tests done, on the effect of channel coding, on the effect of variable CP, and under different SUI Channel Models using the WiMAX PHY layer. The BER curves have been used to compare and analyse the performance of different modulation and coding scheme used. These results provide us quite a comprehensive overview of the WiMAX systems using OFDM physical layer under varying conditions of the wireless channel.

From the performance analysis done, we can conclude the performance of Fixed WiMAX as,

Binary Phase Shift Keying (BPSK) is very power efficient and also requires quite less bandwidth, but has low data rates. However 64-Qadrature Amplitude Modulation (64-QAM) has very high bandwidth requirements but it also provides very high data rates. Quadrature Phase Shift Keying (QPSK) and 16-QAM modulation techniques, on the other hand provide data rates in between BPSK and 64-QAM; they require more bandwidth than BPSK but less than 64-QAM; they also have lower power efficiency then BPSK but better than 64-QAM.

On the case of BER, BPSK provides the least values of BER, while 64-QAM provides the highest values of BER, and QPSK and 16-QAM providing values in between the two, BPSK and 16-QAM.